\documentclass[aps,pra,superscriptaddress,showpacs,twocolumn]{revtex4-1}
\usepackage{graphicx}
\usepackage{amsfonts}
\usepackage{amssymb}
\usepackage{amsmath}
\usepackage[utf8]{inputenc}
\usepackage[T1]{fontenc}
\usepackage{lmodern}

\begin{document}
\newcommand{\ii}{\text{i}}

\title{Bloch dynamics in lattices with long range hoppings}

\author{J. Stockhofe}
\email{jstockho@physnet.uni-hamburg.de}
\affiliation{Zentrum f\"ur Optische Quantentechnologien,
  Universit\"at Hamburg, Luruper Chaussee 149, 22761 Hamburg, Germany}

\author{P. Schmelcher}
\affiliation{Zentrum f\"ur Optische Quantentechnologien,
  Universit\"at Hamburg, Luruper Chaussee 149, 22761 Hamburg, Germany}
\affiliation{The Hamburg Centre for Ultrafast Imaging, Universit\"at Hamburg, Luruper Chaussee 149, 22761 Hamburg, Germany}

\begin{abstract}
We study a discrete Schrödinger equation with arbitrary long range hopping terms under the influence of an external force.
The impact of long range hoppings on the single particle Bloch dynamics in the lattice is investigated.
A closed expression for the propagator is given, based on which we analyze the dynamics of initially Gaussian wave packets.
Our findings capture the anharmonic oscillations recently observed in zigzag lattices and furthermore provide
a detailed quantitative description of the crossover between center of mass Bloch oscillations for wide wave packets and left-right symmetric width oscillations for narrow single site excitations.
The analytical results are shown to be in agreement with numerical simulations.
A helix lattice setup for ultracold atoms is proposed where such hopping terms to far neighbors can be experimentally tuned to sizable values.
\end{abstract}

\maketitle

\section{Introduction}

One of the intriguing features of quantum mechanics in lattice systems is the Bloch oscillation of a particle subjected to a constant external force \cite{Bloch1929,Zener1934}. 
The technical advance of recent years has made it possible to prepare quantum systems clean enough to directly observe this phenomenon,
and Bloch oscillation dynamics have been verified in semiconductor superlattices \cite{Feldmann1992} (see \cite{Rossi1998} for a review), for ultracold atoms subject to optical lattices \cite{Dahan1996,Wilkinson1996}
and in photonic waveguide systems \cite{Morandotti1999,Pertsch1999,Sapienza2003}, among others.
Nowadays, Bloch oscillation based methods are routinely used in cold atom applications, e.g. for precision measurements of the fine structure constant \cite{Clade2006} or gravitational forces \cite{Anderson1998,Roati2004}, even on very small length scales \cite{Ferrari2006}.\\
The discrete Schrödinger lattice model which exhibits Bloch oscillation dynamics under a constant force arises in many different areas of research, both in physics and beyond \cite{Kevrekidis2009}.
Traditionally, this dynamical model has been studied mostly within the nearest neighbor approximation, which is sufficiently accurate in many cases.
Recent years have seen an increased interest in extensions of the model where this approximation is relaxed and long range hoppings (sometimes also termed ``couplings'', depending on the context) are taken into account.
For instance, long range hopping terms have been suggested to be relevant for the dynamics of the DNA molecule \cite{Gaeta1990,Dauxois1991} (recently reviewed in \cite{Zdravkovic2011}) or for excitation transfer in large molecules \cite{Gaididei1997},
superlattice structures with sizable second neighbor hoppings have been proposed \cite{Zhao1997}
and zigzag arrangements of photonic waveguides have been demonstrated as an efficient way to artificially enhance the second neighbor hopping term in a controlled way \cite{Dreisow2008}.
Nonlinear features of this zigzag model have been studied theoretically \cite{Hennig2001,Efremidis2002,Kevrekidis2003} and experimentally \cite{Szameit2009}. 
In \cite{Wang2010} it was proposed that the second neighbor hopping crucially alters the Bloch oscillation dynamics, and this was confirmed in the experiment \cite{Dreisow2011}.
A general theoretical framework that captures all these phenomena has not been given, although crucial building blocks are known.
In \cite{Szameit2008} the propagator for a force-free lattice with arbitrary long range hoppings is derived, and in \cite{Martinez2012} general results for its diffusive dynamics are provided.
Complementing this, the propagator in the presence of a constant force has been derived in \cite{Fukuyama1973}, but restricting to short range hoppings.
There has also been a lot of interest in driven lattice systems (where the force is time-dependent), and also in this context some findings beyond the nearest neighbor approximation are available, e.g. \cite{Dunlap1986,Zhu1999,Jivulescu2006}.\\
In the present work, we extend these results in two ways. First, we derive the propagator for a homogeneous one-dimensional discrete Schrödinger lattice with arbitrary long range hoppings under the influence of a constant external force.
Second, we employ this propagator to systematically study the dynamics of a wave packet depending on its initial width.
Restricting to a nearest neighbor model, a first step in this direction has been done recently in \cite{Dominguez-Adame2010}.
It is well-known (and has been observed in experiments) that for sufficiently wide wave packets the semi-classical picture of Bloch oscillations of the center of mass coordinate applies. In the other extreme limit, where the wave packet is so narrow that it effectively only excites a single site initially, the dynamics is completely different: Under the influence of the force the width of the wave packet oscillates at the Bloch period while its center of mass remains at rest, see e.g. \cite{Grifoni1998}. This periodic wave packet reconstruction goes under different names in the literature, we will refer to it as {\it Bloch breathing} in the following. Below, we will give a detailed analysis of the crossover between Bloch breathing dynamics of initially narrow wave packets and Bloch oscillation dynamics of initially wide ones, taking into account the full set of long range hopping terms.\\
Our study of long range hopping models is motivated by a generalization of the zigzag lattice geometry possible in the framework of cold atoms. In the photonic waveguide systems, one of the three spatial coordinates enters the equation of motion as a time variable \cite{Garanovich2012}. Thus, there are effectively only two spatial degrees of freedom for the lattice layout. In contrast, trapping potentials for cold atoms can be designed in all three spatial dimensions. We suggest a helical arrangement of lattice sites that allows to make arbitrary long range hopping terms (even beyond the second nearest neighbor) sizable. A similar helix model has been briefly considered for fermions in \cite{Wang1991,Xiong1992}.\\
The paper is structured as follows: In Sec.~\ref{sect2} we describe the helix lattice setup that motivates our study of long range hopping models. Sec.~\ref{sect3} introduces the discrete Schrödinger model we consider and lists a number of results for the force-free case. Taking into account the external force, we construct the Wannier-Stark basis and the direct space representation of the propagator in Sec.~\ref{sect4}. Based on this propagator, we study the crossover between Bloch breathing and Bloch oscillations in Sec.~\ref{sect5}. Sec.~\ref{sect6} contains a brief summary and an outlook.

\section{Helix lattice model}
\label{sect2}
\begin{figure}[ht]
                \includegraphics[width=5.0cm]{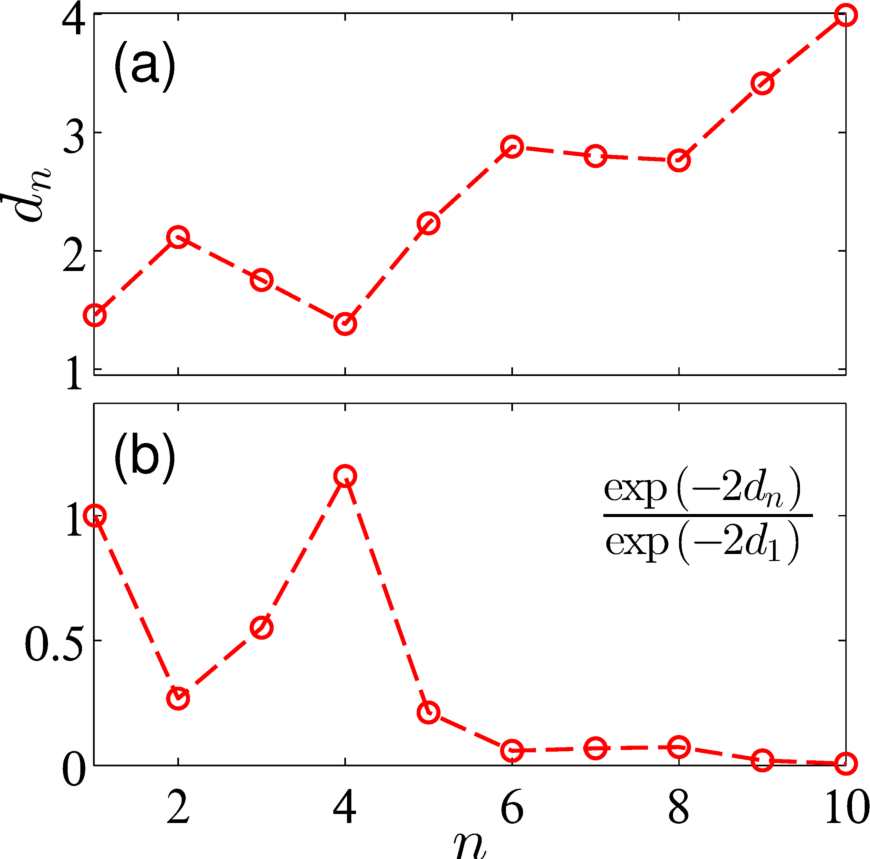}
                \includegraphics[height=4.5cm]{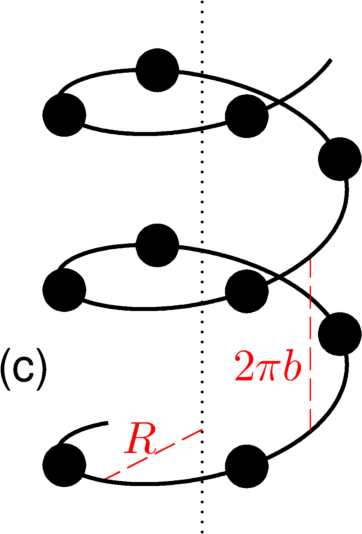}
        \caption{(Color online) (a) Euclidean distances to the $n$-th neighbor site and (b) simple exponential estimate of the scaling of the hopping amplitudes for the helix lattice model schematically shown in (c).}\label{fig:model}
\end{figure}

Zigzag lattice geometries have been successfully used to artificially enhance the second neighbor hopping in planar lattices. In the following we describe a generalization of this idea that allows to enhance also higher order hoppings, at the price of not restricting to planar setups. The scheme we describe relies on the possibility of trapping ultracold atoms on helical space curves, as has been proposed recently \cite{Bhattacharya2007,Reitz2012}.
A single strand helix curve is parametrized by $\vec r (\varphi) = \left(R \cos \varphi,R \sin \varphi, b \varphi \right)$ with $R$ the helix radius and $2 \pi b$ the helix pitch.
Now let us consider potential wells located along the helix equidistantly in arc length (or, equivalently, in the angle parameter $\varphi$), i.e. the $l$-th potential well is centered at 
\begin{equation}
 \vec r_l = \left(R \cos(l \varphi_0), R \sin( l \varphi_0), b \varphi_0 l\right)
\end{equation}
for some fixed $\varphi_0$. Then due to the special features of the helix geometry \cite{Schmelcher2011,Zampetaki2013} the three-dimensional Euclidean distance between sites $j$ and $l$ only depends on the index difference $|j-l|$. Measured in units of the radius $R$, it is given by
\begin{equation}
 d_{|j-l|}^2 = \frac{| \vec r_j - \vec r_l |^2}{R^2} = 2 \left(1-\cos \left[ \varphi_0 (j-l) \right] \right) + \frac{b^2}{R^2} \varphi_0^2 (j-l)^2.
\end{equation}
Interestingly, depending on the helix parameters, $d_n$ can be a non-monotonous function of $n$. This is illustrated in Fig.~\ref{fig:model}~(a) which shows $d_n$ for a helix lattice with parameters $b/R = 0.22$ and $\varphi_0 = \pi/2$. A helix of this geometry is shown in Fig.~\ref{fig:model}~(c), with circular black markers indicating the lattice sites. For this particular geometry, it can be seen that the fourth neighbor (when counted along the helix) of each site is again particularly close to that site in three-dimensional space, although it is far away in index (or in arc length). 

Let us assume now that the dynamics of quantum particles on such a helix lattice can be described by a non-interacting single band Hubbard model of the form \cite{Essler2005,Pethick2008}
\begin{equation}
 \hat H = \sum_{l=-\infty}^\infty \left( -\sum_{\alpha=1}^\infty t_\alpha ( \hat a_l^\dagger \hat a_{l+\alpha} + \hat a_{l+\alpha}^\dagger \hat a_l ) + V_l \hat a_l^\dagger \hat a_l  \right) 
\label{eq:BH}
\end{equation}
where $\hat a_l$ and $\hat a_l^\dagger$ denote annihilation and creation operators (bosonic or fermionic) at site $l$, respectively, $t_\alpha$ is the hopping term to the $\alpha$-th neighbor and $V_l$ models a local scalar potential at site $l$. 
Then the helix geometry will make it necessary to take into account hopping terms with $\alpha > 1$, i.e. beyond nearest neighbor in index. This naturally generalizes the idea of planar zigzag lattices in which the second neighbor hopping can be of the same order of magnitude as the nearest neighbor hopping to the three-dimensional space, opening a way to make even more long range hopping terms sizable. Crucially, we assume that the translational invariance of the model is preserved in the sense that the hopping amplitudes depend on index difference only. This is ensured if the full lattice potential landscape is invariant under the screw operation that for every $l$ maps $\vec r_l$ to $\vec r_{l+1}$. 
We take the atoms to be effectively non-interacting here, a requirement that can experimentally be realized for many species by means of Feshbach resonance techniques \cite{Chin2010}.\\
In the above, we have simply assumed the existence of a Hubbard Hamiltonian as in Eq.~(\ref{eq:BH}). In other words, it is assumed that the three-dimensional potential landscape that constitutes the potential wells is such that Wannier-type modes localized at each lattice site exist and that in the regime considered it is sufficient to restrict to one such mode per site. While the existence and construction of localized Wannier modes for 1D lattices with the usual discrete translation symmetry is well-known, this is much less the case for a system with a basic symmetry under discrete screw operations, such as the helix lattice. Recently, it has been shown on abstract grounds that localized Wannier functions still exist for the latter type of problem, and a tight-binding Hamiltonian as in Eq.~(\ref{eq:BH}) has been justified \cite{Cornean2008}.\\
In the present work, we do not aim at constructing the Wannier states and calculating the Hubbard parameters from first principles. 
For a rough idea of the scaling of the hopping amplitudes $t_\alpha$, we resort to the result of \cite{He2001} where the hoppings in a lattice with the usual discrete translational symmetry are shown to decay exponentially with the Euclidean inter-site distance, up to a polynomial prefactor. 
Let us assume, solely for the purpose of illustration, that this scaling also applies in the helix lattice landscape, say $t_n \sim \exp(-2 d_n)$.
This quantity is shown in Fig.~\ref{fig:model}~(b), normalized to its value at $n=1$. Clearly, there are several values whose order of magnitude is comparable to the $n=1$ contribution.
For the chosen geometry parameters, the dominant hoppings will be $t_1$ (to the nearest sites along the winding) and $t_4$ (to the nearest sites on the adjacent windings), with a strong decay beyond $t_5$.
We emphasize that the actual helix hopping parameters will depend on the detailed shape of the lattice potential, the heights of the potential barriers along and between the windings etc.,
and finding them would require a band structure calculation for the specific potential taking into account the screw symmetry. 
Since this may be a hard task, below we also suggest a way to experimentally extract the parameters from the Bloch dynamics of the system under a constant external force along the $z$-axis, shown as dotted line in Fig.~\ref{fig:model}~(c). Note that the helix geometry is such that for a potential linear in $z$, $V(\vec r)=F_0 z$, the values at the lattice sites scale linearly with the index, i.e. $V (\vec r_l) = F_0 b \varphi_0 l$. \\
Finally, we remark that arguably the proposals presently available suggest that a double helix potential for cold atoms may be easier to realize experimentally than the single helix strand we assume here. We can still restrict to the simple single strand scenario since a non-interacting double helix model with hoppings along and between the strands can be reduced to two decoupled copies of the single strand model, as shown in appendix \ref{app:doublehelix}.


\section{Equation of motion and force-free limit}
\label{sect3}
In the following, we give a detailed analysis of the single particle dynamics governed by the Hamiltonian (\ref{eq:BH}). Expanding the state at time $\tau$ as $|\Psi \rangle_\tau = \sum_l \Psi_l(\tau) \hat a_l^\dagger | \emptyset \rangle$ where the $\Psi_l$ are complex coefficients and $| \emptyset \rangle$ denotes the vacuum state, the Schrödinger equation for $|\Psi \rangle_\tau$ yields an equation of motion for the coefficients which in dimensionless form reads
\begin{equation}
 \ii \partial_\tau \Psi_l(\tau) = - \sum_{\alpha=1}^A t_\alpha \left( \Psi_{l+\alpha}(\tau) + \Psi_{l-\alpha}(\tau) \right) + F l \Psi_l (\tau).
\label{eq:DS}
\end{equation}
This result is obtained both for bosonic and fermionic ladder operators. 
For bosons, the same equation arises when approaching the non-interacting limit from the Gross-Pitaevskii regime \cite{Trombettoni2001}.
We have specified here the potential term $V_l$ to be linear in $l$, modeling a constant force along the $z$-axis in the helix example. 
For technical reasons, we take the number of non-zero hopping terms to be finite in the following, but the cutoff $A$ can be arbitrarily large.
We emphasize that while the helix lattice for ultracold atoms motivates our study of long range hopping terms, the following considerations are based solely on Eq.~(\ref{eq:DS}) and apply in any context in which this equation arises.

\begin{figure}[ht]
\centering
		\includegraphics[width=8.3cm]{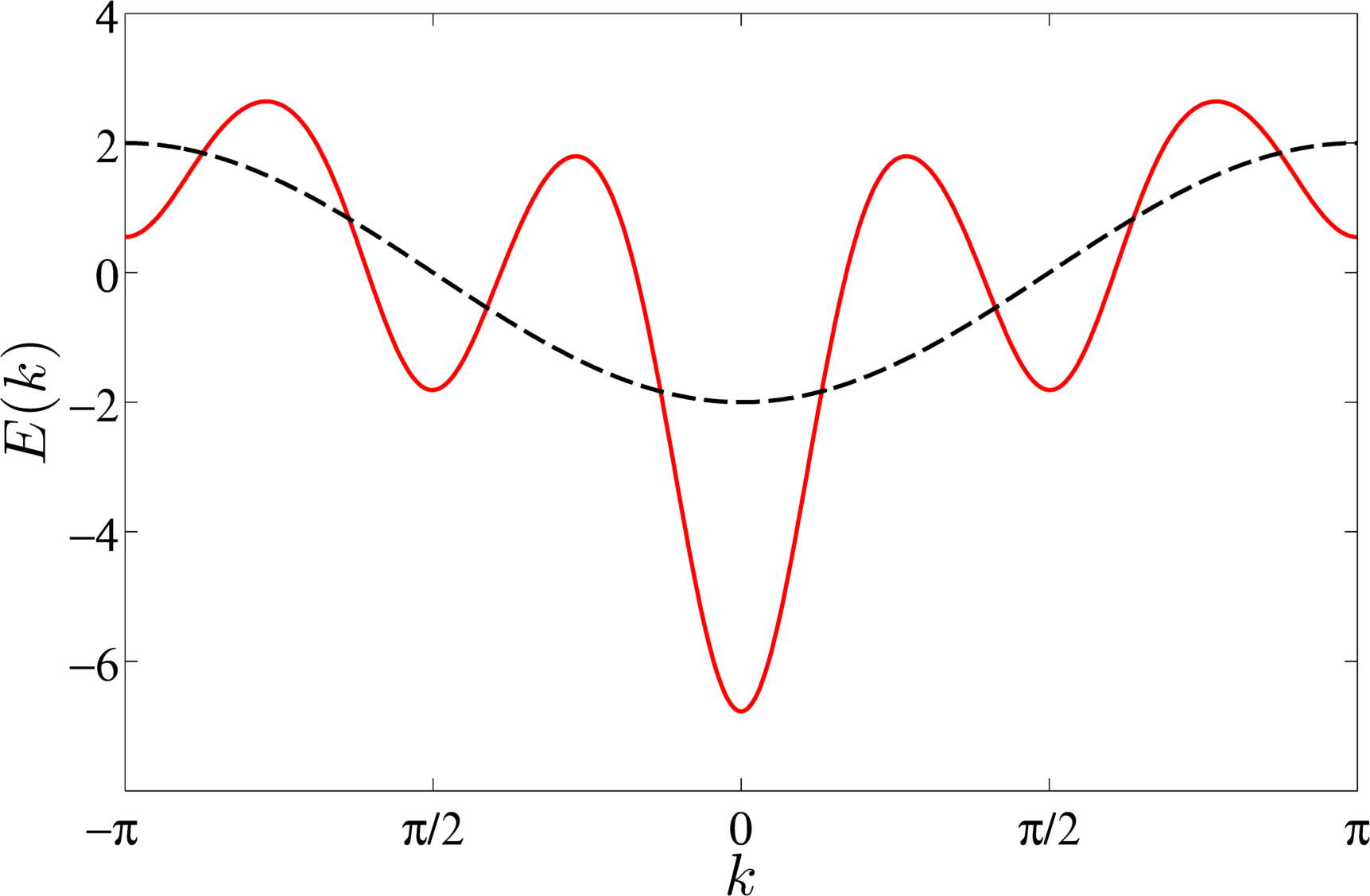}
        \caption{(Color online) Band structure given by Eq.~(\ref{eq:disp}) for a model with long range hoppings as in Fig.~\ref{fig:model} (b) (red solid line) and of a nearest neighbor model with only $t_1=1$ (black dashed line).}\label{fig:bandstruct}
\end{figure}

As a first step, we establish some results on the discrete Schrödinger equation (\ref{eq:DS}) in the force-free limit.
As usual, stationary solutions are obtained by factorizing $\Psi_l(\tau) = \psi_l \exp(-\ii E \tau)$. It is straightforward to see that in the absence of an external force, $F=0$, Eq.~(\ref{eq:DS}) admits stationary plane wave solutions of the form
$\psi_l \propto \exp(\ii k l)$. This is a direct consequence of the translational invariance of the model. The corresponding dispersion relation is given by
\begin{equation}
 E(k) = - 2 \sum_{\alpha=1}^A t_\alpha \cos(k \alpha), \qquad -\pi < k \leq \pi.
\label{eq:disp}
\end{equation}
It is apparent from Eq.~(\ref{eq:disp}) that the presence of long range hopping terms leads to the emergence of the corresponding higher harmonics of $k$ in the dispersion relation. Depending on the relative values of the $t_\alpha$, this can dramatically alter the shape of the $E(k)$ curve, inducing for instance additional maxima and minima or causing a shift of the band edges to different values of $k$. An example of this is shown in Fig.~\ref{fig:bandstruct}, where for definiteness we have used the values displayed in Fig.~\ref{fig:model} (b) for the $t_\alpha$ (with a cutoff at $A=8$).

For the zigzag limiting case of only $t_1$ and $t_2$ non-vanishing, this type of deformation of the band structure has been discussed in detail in \cite{Efremidis2002}, pointing out also a number of consequences for the localized solutions of the corresponding discrete nonlinear Schrödinger model. 
A general property to be read off from Eq.~(\ref{eq:disp}) is that the global minimum of $E$ is assumed at $k=0$ as long as all $t_\alpha \geq 0$. That is, if the signs of all hopping parameters are such that they energetically favor the linked sites to be in phase, the ground state is simply given by the zero quasi-momentum state. If, in contrast, there are hopping terms with $t_\alpha <0$ present in the system, this favors a phase difference of $\pi$ between sites of distance $\alpha$. Depending on the detailed values of the $t_\alpha$, this may cause frustration and thus lead to a non-trivial ground state. Again restricting to the zigzag limit, phase diagrams of interacting models in this interesting frustrated regime have been studied recently in \cite{Greschner2013,Dhar2013}.


\section{Wannier-Stark states and propagator}
\label{sect4}
\begin{figure}[ht]
\centering
                \includegraphics[width=7.6cm]{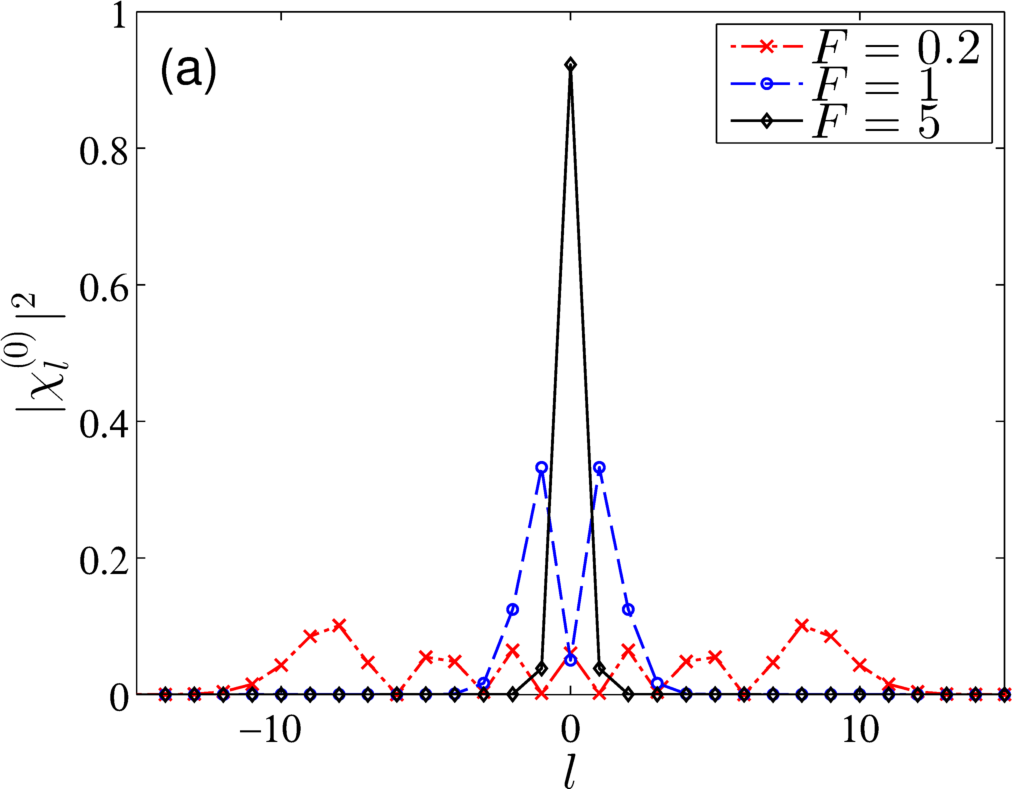}
                \includegraphics[width=7.6cm]{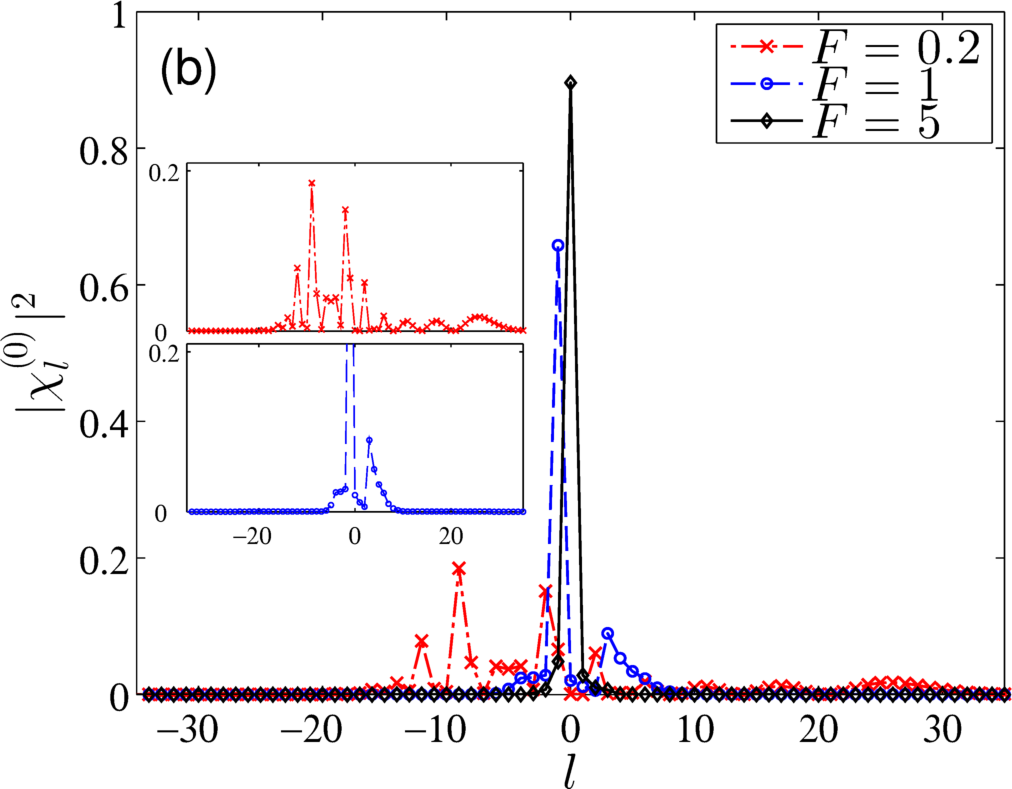}
        \caption{(Color online) Density profiles of Wannier-Stark states centered at $l=0$ at different values of $F$ as listed in the legend for (a) a nearest neighbor model with only $t_1 = 1$, (b) a model with long range hoppings $t_1=t_2=t_3=1$, where the insets show a zoom of the low density region for the less localized states at smaller $F$.}\label{fig:wannierstark}
\end{figure}

As soon as the force parameter $F \neq 0$, plane waves no longer solve Eq.~(\ref{eq:DS}). Instead, there is now a complete set of localized modes, the Wannier-Stark states \cite{Wannier1960,Wannier1962,Fukuyama1973}. Their construction for the discrete Schrödinger model by transforming to momentum space, as detailed for example in \cite{Hartmann2004}, can be extended in a straightforward way to take into account the presence of long range hoppings. We find the Wannier-Stark modes to be given by 
\begin{equation}
 \chi_l^{(w)} = \frac{1}{2\pi} \int_0^{2\pi}\text d k \exp \left[ \ii (l-w)k -\ii \sum_{\alpha=1} ^A \frac{2 t_\alpha}{\alpha F} \sin(k \alpha) \right].
\label{eq:ws1}
\end{equation}
Each Wannier-Stark state is labeled by an integer quantum number $w \in \mathbb Z$, and its energy eigenvalue is given by $E_w = w F$. This equidistant spectrum, the so-called Wannier-Stark ladder, is insensitive to the presence of long range hoppings. The fact that all energy eigenvalues are harmonics of the fundamental frequency $F$ immediately implies that the dynamics of any initial state is periodic with period $T_B = 2 \pi/F$. This includes the famous Bloch oscillations of a wave packet localized in momentum space, but is also true for any other initial condition.\\
To proceed, we use the Jacobi–Anger expansion \cite{Gradshteyn2007}
\begin{equation}
 e^{-iz \sin \phi} = \sum_{n=-\infty}^\infty J_n(z) e^{-in\phi},
\end{equation}
where $J_n$ denotes the $n$-th Bessel function of the first kind. Then the integral in Eq.~(\ref{eq:ws1}) can be evaluated to
\begin{equation}
 \chi_l^{(w)} = \sum_{ n_1, \dots, n_A } \delta_{\sum_{\alpha=1}^A \alpha n_\alpha, l-w} \prod_{\alpha=1}^A J_{n_\alpha}\left( \frac{2 t_\alpha}{\alpha F} \right),
\label{eq:ws2}
\end{equation}
where all indices $n_\alpha$ are summed over $\mathbb Z$.
In the limit of only nearest neighbor hopping, $t_\alpha = 0$ for $\alpha \geq 2$, we have $J_{n_\alpha}\left( \frac{2 t_\alpha}{\alpha F} \right) = \delta_{n_\alpha,0}$ for $\alpha \geq 2$, such that $\chi_l^{(w)} = J_{l-w}\left( \frac{2 t_1}{F} \right)$ in this limit, reproducing the result of \cite{Fukuyama1973}.\\
Using basic properties of the Bessel functions (see also below), we can immediately establish some properties of the Wannier-Stark states from Eq.~(\ref{eq:ws2}). First, the Wannier-Stark mode of quantum number $w$ is centered at site $w$, since $\sum_l l |\chi_l^{(w)}|^2 = w$. Second, for increasing values of $F$ the mode becomes more and more localized, since 
\begin{equation}
 \sum_l l^2 |\chi_l^{(w)}|^2 - w^2= \frac{2}{F^2} \sum_{\alpha=1}^A t_\alpha^2. 
\end{equation}
We see here how the presence of long range hopping terms favors a delocalization of the Wannier-Stark states. Single site localization sets in when the force is so strong that $F^2 \gg \sum_\alpha t_\alpha^2$.\\
When restricting to next neighbor hopping, the Wannier-Stark mode has the symmetry $\chi^{(w)}_{w-l} = (-1)^l \chi^{(w)}_{w+l}$ which ensures that the density is symmetric around the central site $w$. This symmetry property is lost immediately if additional hoppings are taken into account. Instead, the stationary localized modes described by Eq.~(\ref{eq:ws2}) generally exhibit intricate asymmetric density distributions around the central site. This is illustrated in Fig.~\ref{fig:wannierstark}, which also shows the localization that increases with $F$.

Since the Hamiltonian is diagonal in the Wannier-Stark basis, we can immediately write down the matrix elements of the propagator
 \begin{equation}
 U_{l, l'}(\tau)=  \sum_{w \in \mathbb Z}  \chi_l^{(w)} \left(\chi_{l'}^{(w)}\right)^* e^{-\ii w F \tau}.
\end{equation}
Inserting the expansion of the Wannier-Stark modes given in Eq.~(\ref{eq:ws2}) and repeatedly making use of the identity \cite{Gradshteyn2007}
\begin{equation}
 \sum_{n=-\infty}^\infty J_n(z)J_{n+j}(z)e^{\ii n \phi} = J_j\left(2 z \sin \frac{\phi}{2} \right) \ii^j e^{-\ii  j \phi/2},
\label{eq:bessel}
\end{equation}
this is evaluated to be
 \begin{eqnarray}
 U_{l, l'}(\tau)&=&  \ii^{l-l'}e^{-\ii \frac{F \tau}{2}(l+l')} \nonumber \\
&\times& \sum_{n_2, \dots, n_A} J_{l-l'-\sum_{\alpha=2}^A \alpha n_\alpha} \left(\frac{4 t_1}{F} \sin \frac{F \tau}{2} \right)  \nonumber \\
&\times& \prod_{\alpha=2}^A  \ii^{-n_\alpha(\alpha-1)} J_{n_\alpha} \left( \frac{4 t_\alpha}{\alpha F} \sin \frac{F \tau \alpha}{2} \right).
\label{eq:prop}
\end{eqnarray}
Again, let us discuss some limiting cases. If all hoppings are turned off, the propagator collapses to $U_{l,l'}(\tau) = \delta_{l,l'} e^{-\ii F l \tau}$, corresponding to decoupled sites whose phases evolve with the respective local potential. When we restrict ourselves to the nearest neighbor hopping term, we arrive at the result of \cite{Fukuyama1973}, see also \cite{Holthaus1996,Grifoni1998,Hartmann2004,Szameit2007b}. Finally, taking the limit of $F \rightarrow 0$ and noting that $\lim_{\epsilon \rightarrow 0} \frac{\sin \epsilon}{\epsilon} = 1$, we recover the propagator of a force-free lattice with arbitrary long range hoppings recently obtained in \cite{Szameit2008}.


\section{Bloch dynamics of wave packets}
\label{sect5}
Having derived the closed expression for the propagator, we are now in the position to study the dynamics of an arbitrary initial wave function. 
Let us start by discussing the simple case of an initial single site excitation, $\Psi_l (\tau=0) = \delta_{l,l_0}$.
Then the wave function at time $\tau$ is obtained using the propagator from Eq.~(\ref{eq:prop}) as
\begin{equation}
 \Psi_l(\tau) = \sum_{l'} U_{l,l'}(\tau) \Psi_{l'} (\tau=0) = U_{l,l_0} (\tau).
\end{equation}
From the properties of the Bessel function, one finds that even in the presence of long range hoppings the propagator has the symmetry $|U_{l_0-l,l_0}(\tau)| =| U_{l_0+l,l_0}(\tau)|$ for any $l_0, l$ and any $\tau$. Thus, during the time evolution following the initial single site excitation, the density will remain left-right symmetric around the central site $l_0$. In particular, the center of mass remains at rest at $l_0$. The width, however, can change as a function of time, and from the equidistant spacing of the Wannier-Stark spectrum it is clear that the dynamics of the width has to be oscillatory with period $T_B = 2\pi /F$. This Bloch breathing behavior is well-known for nearest neighbor models.

On the other hand, if the initial excitation is not restricted to a single site, but rather a wave packet that is sufficiently delocalized in position space to be considered localized in $k$-space, then this wave packet will perform Bloch oscillations under the influence of the external force. The temporal periodicity is the same as for the Bloch breathings, but now the center of mass of the wave packet performs oscillations while its shape is essentially unchanged. In most cases the theoretical description of these Bloch oscillations is based on a semi-classical argument explicated below.

In spite of the qualitative differences between Bloch breathing and Bloch oscillations, these dynamics are two sides of the same coin and are both captured by the propagator in Eq.~(\ref{eq:prop}).
In the following, our aim is two-fold: First, we discuss the crossover from Bloch breathing to Bloch oscillations by considering an initial Gaussian wave packet of variable width. 
This is done for arbitrary long range hoppings, and in a second step we work out the limiting cases of pure Bloch breathing and pure Bloch oscillations for lattices with long range hoppings
and provide quantitative estimates for which parameter values of the model these limits apply.

To explore the crossover region, we consider as the initial state a site-centered Gaussian wave packet defined by
\begin{equation}
 \Psi_l (\tau=0) = \mathcal{N} e^{-l^2/(2 \sigma)}, \qquad \sigma > 0.
\label{eq:initial}
\end{equation}
More general initial conditions are discussed in appendix \ref{app:prop}.
Without loss of generality we take the wave packet to be centered at site $0$ initially. In the limit of small $\sigma$, Eq.~(\ref{eq:initial}) describes essentially a single site excitation that, as seen above, will perform Bloch breathing. In contrast, if $\sigma$ is large we expect Bloch oscillations of the wave packet's center of mass, with no notable deformation.
\\
First of all we find that the normalization constant $\mathcal N$ is evaluated to be $1/ \mathcal N^2 = \sum_l e^{-l^2/\sigma}  = \vartheta_3 \left(0, e^{-1/\sigma} \right)$,
where $\vartheta_3$ denotes the Jacobi Theta function of the third kind \cite{Bellmann1961} defined as
\begin{equation}
 \vartheta_3 (u,q) = 1 + 2 \sum_{n=1}^\infty q^{n^2} \cos 2 n u, \qquad | q | < 1,
\end{equation}
which will also become important below.  
With this, we can evaluate the two key quantities to characterize and distinguish Bloch breathing and Bloch oscillations, namely the first two moments of the position operator, 
$\langle X \rangle_\tau = \sum_l l |\Psi_l(\tau)|^2$ and $\langle X^2\rangle_\tau = \sum_l l^2 |\Psi_l(\tau)|^2$.
These quantities can be obtained through a lengthy but direct calculation, making repeated use of the Bessel function orthogonality and recursion identities \cite{Gradshteyn2007} (the former being a special case of Eq.~(\ref{eq:bessel}))
\begin{subequations}
 \begin{eqnarray}
 && \sum_l J_l(z) J_{l+j}(z) = \delta_{j,0}, \\
 && l J_l(z) = \frac{z}{2} \left( J_{l-1}(z) + J_{l+1}(z) \right).
\end{eqnarray}
\label{eq:bessel2}
\end{subequations}

\begin{figure}[ht]
                  \includegraphics[width=8cm]{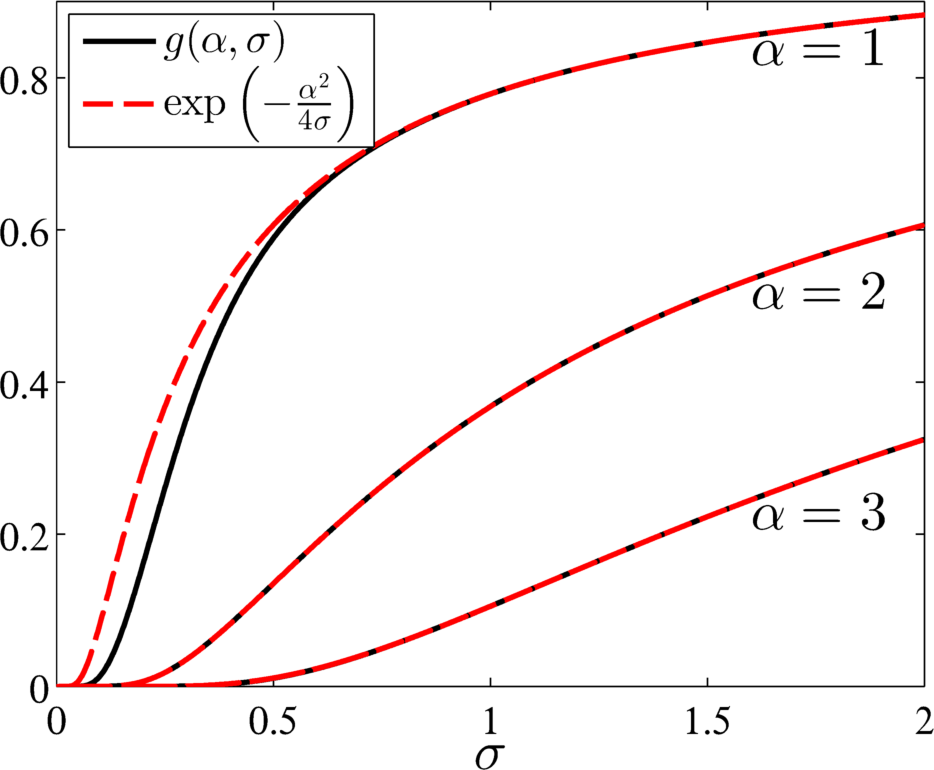} 
        \caption{(Color online) Full crossover function $g$, compared to the plain exponential factor, as a function of $\sigma$ for various values of $\alpha$. In the limit of large $\sigma$, all curves tend to $1$.}\label{fig:plotg}
\end{figure}

Defining the auxiliary function $g(\alpha,\sigma)$ for $\alpha \in \mathbb{N}$ as 
\begin{eqnarray}
  g(\alpha, \sigma) &=& \sum_l \Psi_l(\tau=0) \Psi_{l+\alpha} (\tau=0) \nonumber\\
 &=& e^{-\alpha^2/(4\sigma)} \,\, \frac{\vartheta_3 \left( \alpha \frac{\pi}{2}, e^{-\sigma \pi^2} \right)}{\vartheta_3 \left( 0, e^{-\sigma \pi^2} \right)},
\end{eqnarray}
we finally obtain compact expressions for the first two moments of the position operator:
\begin{widetext}
\begin{eqnarray}
\langle X \rangle_\tau &=& - \sum_{\alpha=1}^A \frac{4 t_\alpha}{F} g \left( \alpha,\sigma \right) \sin^2 \frac{\alpha F \tau}{2} , \label{eq:x}\\
\langle X^2 \rangle_\tau &=&  \langle X^2 \rangle_{\tau=0} + \sum_{\alpha=1}^A \frac{8 t_\alpha^2}{F^2} \sin^2 \frac{\alpha F \tau}{2}\left[ 1 - g \left( 2\alpha,\sigma \right) \cos (\alpha F \tau)   \right] \nonumber \\
&&+\sum_{\alpha > \beta} \frac{16 t_\alpha t_\beta}{F^2} \sin \frac{\alpha F \tau}{2} \sin \frac{\beta F \tau}{2} \left[ g \left( \alpha -\beta,\sigma \right)  \cos \frac{(\alpha -\beta)F \tau}{2}  
-g \left( \alpha+\beta,\sigma \right) \cos \frac{(\alpha +\beta)F \tau}{2}  \right] \label{eq:x2}.
\end{eqnarray}
\end{widetext}
Details on the derivation of Eqs.~(\ref{eq:x},\ref{eq:x2}) are provided in appendix \ref{app:prop}, where also the corresponding generalizations to arbitrary initial conditions are given. 
From these equations, it is immediately seen that the presence of long range hoppings introduces higher harmonics of the fundamental Bloch frequency $F$ in the time evolution of $\langle X \rangle_\tau$ and $\langle X^2 \rangle_\tau$.
Intuitively, these harmonics are expected to show up since the potential difference between two sites of index difference $\alpha$ is proportional to $\alpha$ in our model, 
thus the effective Bloch frequency when considering hopping only to the $\alpha$-th neighbor is given by the harmonic $\alpha F$.
The second observation to be made is that $g$ acts as a crossover function, interpolating between Bloch breathing for small $\sigma$ and Bloch oscillations for large $\sigma$. 
This is reflected by the limits
\begin{equation}
 g(\alpha, \sigma) \longrightarrow \quad \begin{cases} 0, \qquad \sigma \rightarrow 0 \\ 1, \qquad \sigma \rightarrow \infty  \end{cases},
\end{equation}
holding for any nonzero $\alpha$. We furthermore see that the overall shape of $g$ is strongly determined by the exponential prefactor. Distinguishing the case of the integer argument $\alpha$ being even or odd, one finds
\begin{equation}
  g(\alpha, \sigma) =  \begin{cases} e^{- \alpha^2/(4\sigma)} \phantom{\,\, \frac{\vartheta_3 \left( \frac{\pi}{2}, e^{-\sigma \pi^2} \right)}{\vartheta_3 \left( 0, e^{-\sigma \pi^2} \right)}}, \qquad \alpha \text{ even}
\\  e^{- \alpha^2/(4\sigma)} \,\, \frac{\vartheta_3 \left( \frac{\pi}{2}, e^{-\sigma \pi^2} \right)}{\vartheta_3 \left( 0, e^{-\sigma \pi^2} \right)}, \qquad \alpha \text{ odd}  \end{cases}.
\end{equation}
The correction factor $\vartheta_3 ( \pi/2, e^{-\sigma \pi^2} )/\vartheta_3 ( 0, e^{-\sigma \pi^2} )$ occuring for $\alpha$ odd here is close to unity for not too small values of $\sigma$, since $\exp(-\sigma \pi^2)$ quickly goes to zero with increasing $\sigma$. Thus, the second argument of the Jacobi Theta functions quickly decays to zero and both Theta functions approach unity. On the other hand, for small $\sigma$ the correction factor goes to zero, so the relative deviation between $g$ and the plain exponential $\exp(-\alpha^2/(4\sigma))$ becomes large. The absolute error, however, is small still, since the value of the exponential itself approaches zero in this limit (the larger $\alpha$, the faster). 
Fig.~\ref{fig:plotg} shows the full $g(\alpha,\sigma)$ for $\alpha=1,2,3$ and in comparison also the exponential factor $\exp(-\alpha^2/(4\sigma))$. For large $\sigma$, where it eventually approaches unity, $g$ 
is fully dominated by the exponential. This also implies that the convergence to $1$ is slower the larger $\alpha$, a 
trend that is clearly observed when going from $\alpha=1$ to $\alpha=3$ already. For small $\sigma$, a notable deviation between $g$ and the exponential can be seen for $\alpha=1$. For $\alpha=2$ (and any other even $\alpha$) the agreement is exact. For $\alpha=3$, the absolute deviations due to the Jacobi Theta function correction factor are already so small that they cannot be identified on the scale of this figure. 

\begin{figure}[ht]
        \centering
		\includegraphics[width=8.5cm]{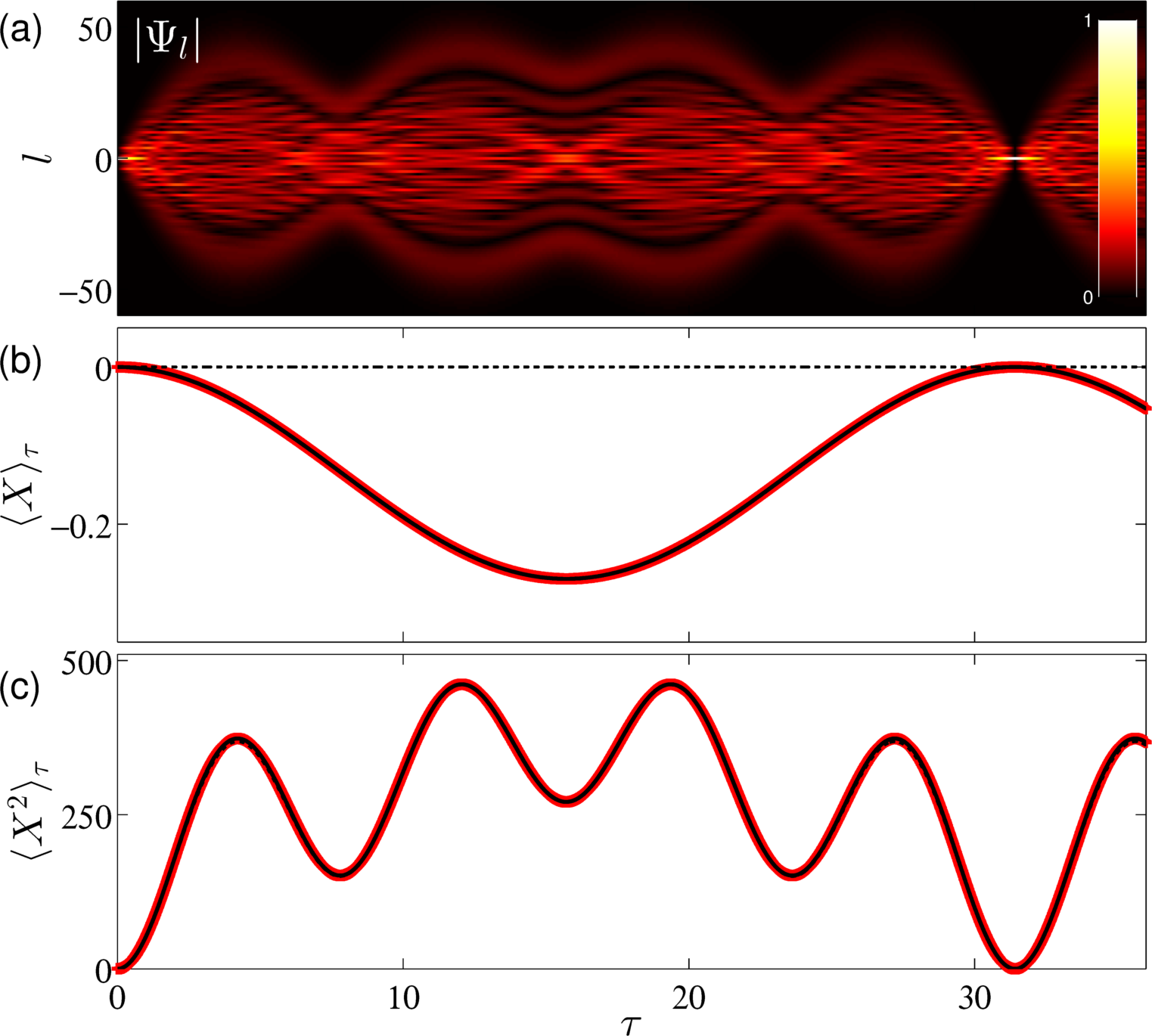}
        \caption{(Color online) Multi-frequency Bloch breathing oscillation of a narrow Gaussian wave packet with $\sigma=0.1$ at $F=0.2$, hopping parameters as in Fig.~\ref{fig:model}(b). (a) Absolute value of $\Psi_l$ as a function of time $\tau$ and site index $l$. (b,c) First two moments of the position operator: numerical values (black lines), analytical result of Eqs.~(\ref{eq:x},\ref{eq:x2}) (red markers) and approximations of Eqs.~(\ref{eq:breathingx},\ref{eq:breathingx2}) (dashed black lines).}\label{fig:sim1}
\end{figure}

Let us now discuss separately the two limiting cases of small and large width of the initial wave packet, respectively.
For $\sigma \rightarrow 0$, so all occurrences of $g \rightarrow 0$ in Eqs.~(\ref{eq:x},\ref{eq:x2}), we find
\begin{eqnarray}
  \langle X \rangle_\tau &=& 0, \label{eq:breathingx} \\
\langle X^2 \rangle_\tau &=&  
\sum_\alpha \frac{4 t_\alpha^2}{F^2} \left[1-\cos (\alpha F \tau ) \right].
 \label{eq:breathingx2}
\end{eqnarray}
We see that in this limit the center of mass is at rest while the wave packet width performs oscillations.
This is characteristic of a pure Bloch breathing dynamics, where the long range hopping terms contribute oscillations at higher harmonics of the Bloch frequency $F$. This includes the two-frequency Bloch breathing recently observed in zigzag lattices with non-vanishing $t_1$ and $t_2$ \cite{Dreisow2011}. Fig.~\ref{fig:sim1} shows the time evolution of a narrow Gaussian wave packet obtained by directly integrating the discrete Schrödinger equation (\ref{eq:DS}). The moments of the position operator are evaluated numerically (red markers) and as a check compared to the analytical expressions of Eqs.~(\ref{eq:x},\ref{eq:x2}) (black lines), showing agreement. The approximate predictions of Eqs.~(\ref{eq:breathingx},\ref{eq:breathingx2}) are shown as dashed lines. It is clearly observed that this situation is very close to the pure Bloch breathing limit, with the center of mass being almost at rest. For the width oscillation, the deviation from Eq.~(\ref{eq:breathingx2}) is almost 
indiscernible on the scale of this figure. The Bloch period is given by $T_B = 10 \pi$ for the chosen parameters. After this time, a full refocusing of the localized excitation is observed.\\
Again, we note that Eq.~(\ref{eq:breathingx2}) contains special cases that have been reported before. The corresponding expression for next neighbor hopping only is given in \cite{Grifoni1998}.
In the force free limit, the recent result of \cite{Martinez2012} for the ballistic spreading of a single site excitation in a lattice with long range hoppings is recovered, i.e. $\langle X^2 \rangle_\tau \rightarrow 2 \tau^2 \sum_\alpha \alpha^2 t_\alpha^2$ for $F \rightarrow 0$. 
Conversely, in the work \cite{Dunlap1986} that is primarily concerned with AC driven nearest neighbor lattice systems, an extension to long range hoppings is sketched which when worked out \cite{Zhao1991} reduces to Eq.~(\ref{eq:breathingx2}) in the static limit.

\begin{figure}[ht]
        \centering
                 \includegraphics[width=8.5cm]{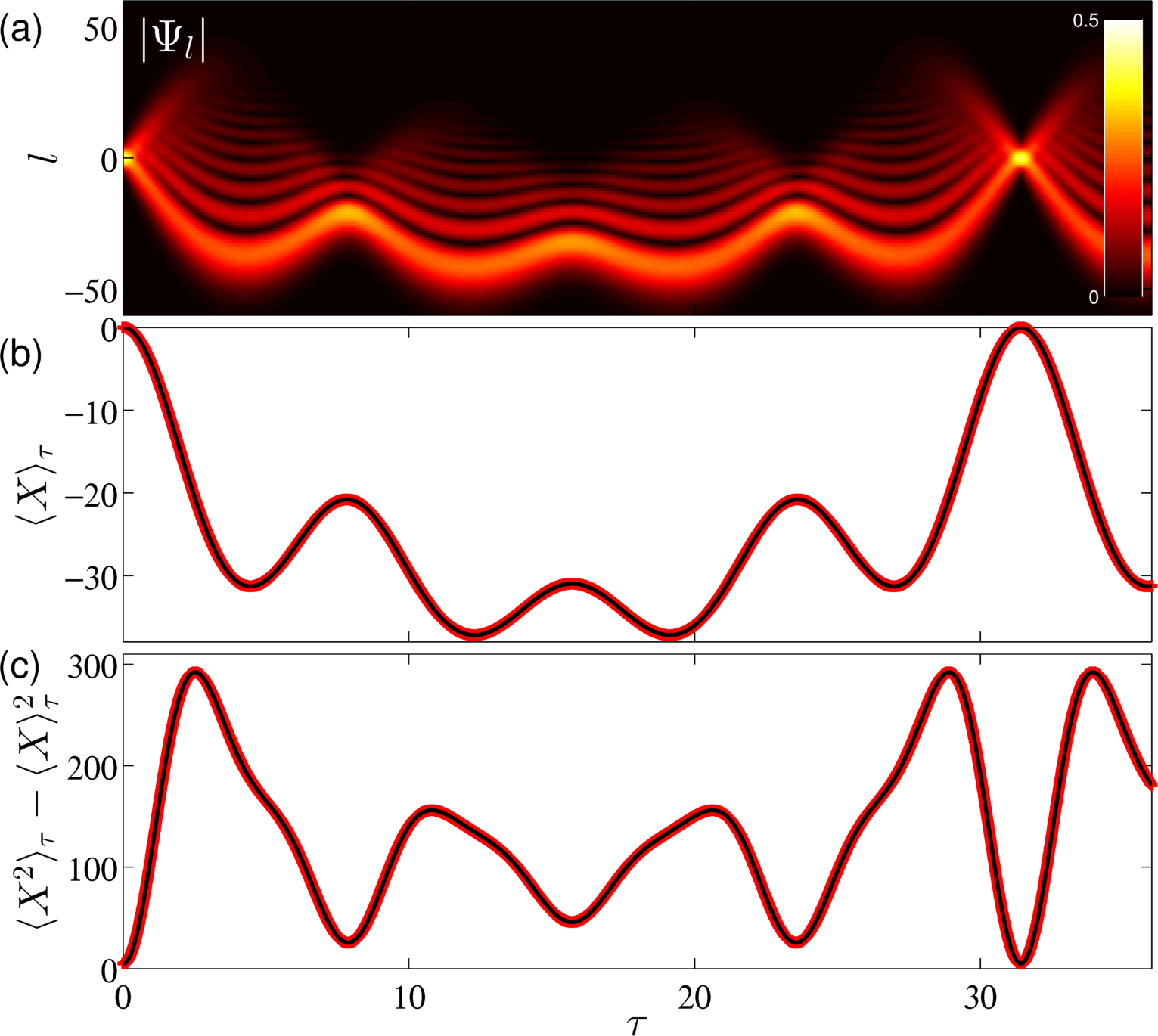}
        \caption{(Color online) Intermediate case with nontrivial dynamics both in center of mass and width for an initial Gaussian wave packet with $\sigma=10$ at $F=0.2$, hopping parameters as in Fig.~\ref{fig:model} (b). (a) Absolute value of $\Psi_l$ as a function of time $\tau$ and site index $l$. (b,c) Expectation value and variance of the position operator: numerical values (black lines) and analytical result of Eqs.~(\ref{eq:x},\ref{eq:x2}) (red markers).}\label{fig:sim3}
\end{figure}

In the opposite limit of $\sigma \rightarrow \infty$, all occurences of $g \rightarrow 1$. Then Eqs.~(\ref{eq:breathingx},\ref{eq:breathingx2}) give
\begin{eqnarray}
  \langle X \rangle_\tau &=& - \sum_\alpha \frac{2 t_\alpha}{F} \left[1 - \cos(\alpha F \tau) \right],\label{eq:blochx}\\
  \langle X^2\rangle_\tau- \langle X \rangle_\tau^2 &=& \langle X^2 \rangle_{\tau=0} = \frac{\sigma}{2}.
\label{eq:blochx2}
\end{eqnarray}
These trajectories describe a multi-frequency Bloch oscillation of the wave packet as a whole, as can be inferred from the time-independent variance of $X$. For two non-vanishing hopping terms, such a two-frequency oscillations has been predicted and observed in \cite{Wang2010,Dreisow2011}. 
\begin{figure}[ht]
        \centering
                 \includegraphics[width=8.5cm]{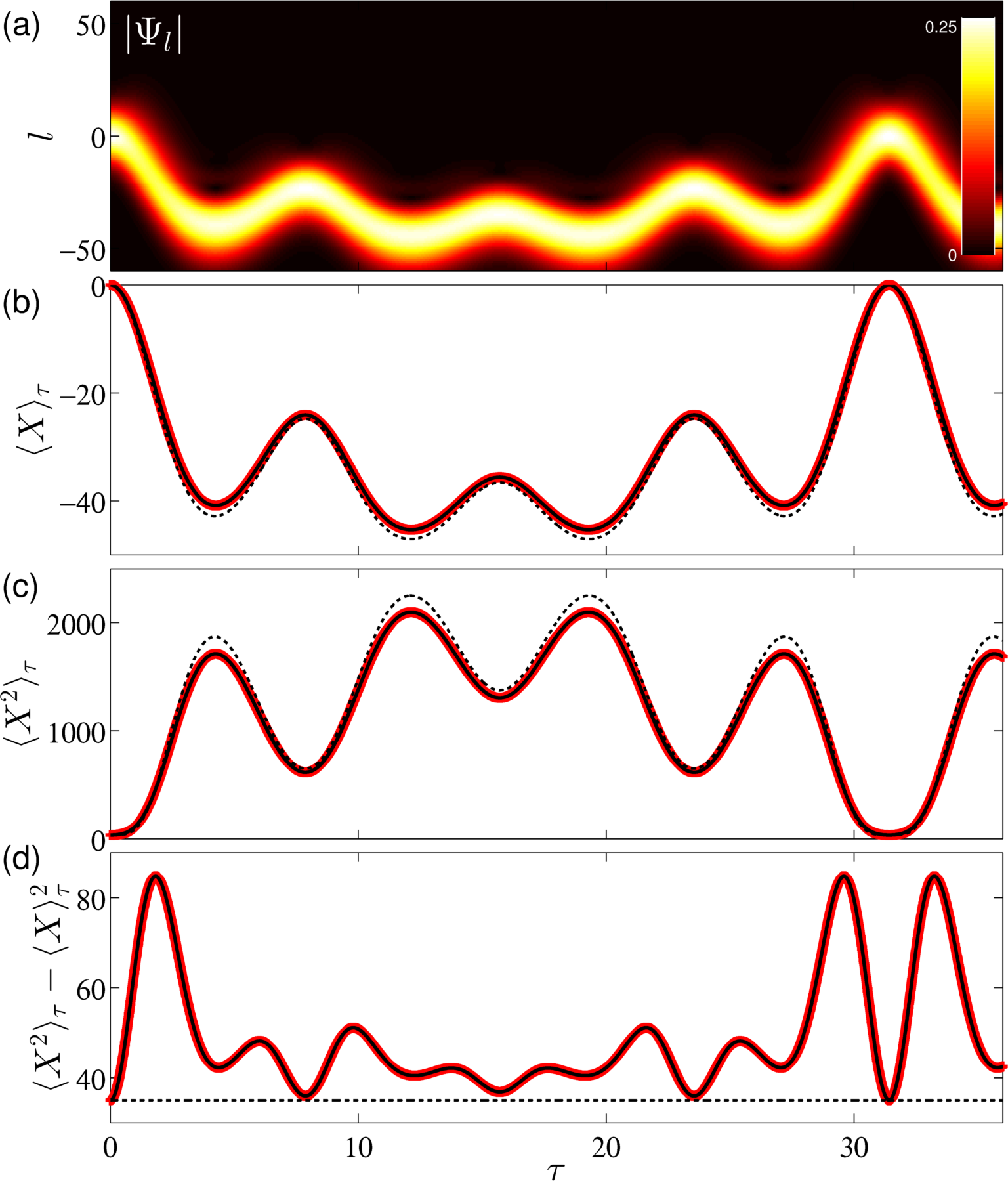}

        \caption{(Color online) Multi-frequency Bloch oscillation of a wide Gaussian wave packet with $\sigma=70$ at $F=0.2$, hopping parameters as in Fig.~\ref{fig:model}(b). (a) Absolute value of $\Psi_l$ as a function of time $\tau$ and site index $l$. (b-d) First two moments and variance of the position operator: numerical values (black lines), analytical result of Eqs.~(\ref{eq:x},\ref{eq:x2}) (red markers) and approximations of Eqs.~(\ref{eq:blochx},\ref{eq:blochx2}) (dashed black lines).}\label{fig:sim2}
\end{figure}
We show now that Eq.~(\ref{eq:blochx}) agrees with the result found from the simple semi-classical argument usually invoked to discuss Bloch oscillations: 
If the initial wave packet is sufficiently delocalized in position space, and thus localized in quasi-momentum space, one can 
obtain the time evolution of its quasi-momentum from the acceleration theorem \cite{Kittel1996,Nenciu1991} as $k(\tau) = - F \tau$. The velocity $v(\tau)$ of the wave packet is then 
given by the group velocity evaluated at $k(\tau)$, i.e.
\begin{equation}
 v(\tau) =\frac{\text d E}{\text d k} \bigg|_{k=-F \tau} = -2 \sum_\alpha \alpha t_\alpha \sin (\alpha F \tau),
\label{eq:v}
\end{equation}
and integrating this equation gives the center of mass motion $\langle X \rangle_\tau$ found in Eq.~(\ref{eq:blochx}).
This way of reasoning directly links the Bloch oscillation trajectory to the band structure. Thus, the additional maxima and minima of $E(k)$ that exist due to the presence of long range hoppings immediately cause additional reversal points of the wave packet dynamics within one Bloch period. In Fig.~\ref{fig:sim2} we present a numerical simulation of a multi-frequency Bloch oscillation, verifying again the validity of Eqs.~(\ref{eq:x},\ref{eq:x2}) against the direct solution of the equation of motion (\ref{eq:DS}).
Indeed, we see that for this comparably large initial width we approach a regime in which the trajectory closely follows the predictions of Eqs.~(\ref{eq:blochx},\ref{eq:blochx2}) shown as dashed lines, although there are still deviations in both $\langle X \rangle_\tau$ and $\langle X^2 \rangle_\tau$. These deviations are particularly visible when calculating the variance that deviates still quite strongly from the constant value $\sigma/2$. We find that going to larger values of $\sigma$, these features are further diminished and the semi-classical equations become more and more exact.\\
Eq.~(\ref{eq:breathingx2}) and Eq.~(\ref{eq:blochx}) allow to calculate the Bloch breathing and oscillation trajectories given the model parameters $t_\alpha$ and $F$. 
Conversely, measuring these trajectories and Fourier transforming them provides a way to extract the hopping parameters $t_\alpha$ and the force $F$ from the time evolution data. 
The frequency spectra will have peaks at the harmonics $\alpha F$ with amplitudes determined by the ratios $t_\alpha/F$. 
We remark that since sparse frequency spectra are expected here, this problem should be an ideal candidate for so-called compressed sensing techniques which allow a reliable reconstruction of the spectrum even from imperfect time signals, see \cite{Candes2008} for an introduction and \cite{Andrade2012} for a recent application in the context of molecular dynamics. \\
So far we have focused on the limiting cases of ideal Bloch breathing on the one hand and ideal Bloch oscillations on the other hand. Eqs.~(\ref{eq:x},\ref{eq:x2}) hold for the full crossover between these two limits, however, and from a study of the function $g$ we can also estimate that there is a large intermediate regime that is neither characterized by a pure breathing nor by a pure center-of-mass oscillation. To reach the limit of Bloch breathing with $\langle X \rangle_\tau \approx 0$, from Eq.~(\ref{eq:x}) we can read off the necessary condition that $g(\alpha,\sigma) \approx 0$ for all $\alpha$ for which $t_\alpha$ is non-negligible. Since $g$ approaches zero faster for $\sigma \rightarrow 0$ the larger $\alpha$, this condition is most restrictive for $\alpha=1$. Consequently, in a system with non-negligible next neighbor hopping we can expect to reach the pure Bloch breathing limit when $g(1,\sigma) \approx 0$, which is the case for $\sigma \lesssim 0.1$ (see Fig.~\ref{fig:plotg}). At $\sigma = 0.
1$, the initial probability to find the particle away from the central site $0$ is already of the order of $10^{-4}$, so to observe pure Bloch breathing one indeed needs an essentially pure single site excitation. Any sizable excitation of neighboring sites will induce deviations from $\langle X \rangle_\tau = 0$ and cause center of mass oscillations according to Eq.~(\ref{eq:x}). In Fig.~\ref{fig:sim1}, it is clearly observed that even at $\sigma = 0.1$ the center of mass motion is not completely suppressed.\\
Turning to the pure Bloch oscillation limit of $\sigma \rightarrow \infty$, from Eq.~(\ref{eq:x}) we can infer the condition $g(\alpha,\sigma) \approx 1$ for all $\alpha$ for which the hopping term $t_\alpha$ is non-negligible. 
We have seen above that the behavior of $g$ at large $\sigma$ is entirely determined by the exponential $\exp(-\alpha^2/(4 \sigma)) \sim 1 - \alpha^2/(4 \sigma)$ for large $\sigma$. Thus, for the center of mass to be described by the semi-classical Bloch trajectory, the condition $4 \sigma \gg \alpha^2$ needs to be satisfied for any $\alpha$ with sizable hopping $t_\alpha$.
From the above semi-classical discussion it is to be expected that the hopping terms of largest $\alpha$ give the strictest condition here: The considerations leading to Eq.~(\ref{eq:v}) rely on a linearization of the dispersion relation $E$ around the central momentum of the wave packet. Higher order hoppings lead to short length scale oscillations in the $E(k)$ curve as seen in Fig.~\ref{fig:bandstruct}, deteriorating the linearization approximation and thus enforcing a stronger localization of the wave packet in $k$-space for the analysis to apply, which in turn demands a larger spread in direct space. Even if $4 \sigma \gg \alpha^2$ is fulfilled for any $\alpha$ with non-negligible $t_\alpha$, it is not directly ensured that the variance also follows the semi-classical expectation $\langle X^2 \rangle_\tau - \langle X \rangle_\tau^2= \langle X^2 \rangle_{\tau = 0}$. For this to hold, we also need the $g \approx 1$ limit to hold for all terms in Eq.~(\ref{eq:x2}). In particular, this poses the necessary 
condition that also $g(2\alpha,\sigma) \approx 1$ for all $\alpha$ with non-negligible $t_\alpha$, implying the stricter condition $\sigma \gg \alpha^2$. In line with this argument, a close inspection of Fig.~\ref{fig:sim2} indeed shows that the relative deviations of the true trajectory 
from the semi-classical expectation are larger for $\langle X^2 \rangle_\tau$ than for $\langle X \rangle_\tau$.\\
Moving away from either of the two limits, the dynamics performed by a Gaussian wave packet of intermediate width consists of nontrivial oscillations both in the center of mass and the variance. An example illustrating this is shown in Fig.~\ref{fig:sim3}.
The initial stage of the dynamics is characterized by a widening of the wave packet into both directions of the lattice, but in an asymmetric way, with a larger fraction moving to $l<0$ and following a Bloch oscillation-type trajectory there. After the Bloch period, the wave packet is fully reconstructed. 

\section{Brief summary and perspectives}
\label{sect6}
We have investigated a translationally invariant discrete Schrödinger equation with arbitrary long range hopping terms. The study of this equation was motivated by a proposal for a helix lattice for ultracold atoms, extending the idea of a planar zigzag lattice to three spatial dimensions. With such a lattice setup it will be possible to tune the long range hoppings of a cold atom Hubbard model through the helix geometry parameters. We have discussed the plane wave solutions and dispersion relation of the force-free model and then turned to the influence of an external force. Closed expressions for the Wannier-Stark states and the propagator were constructed in terms of Bessel functions. Finally, we analyzed the Bloch dynamics of a Gaussian wave packet of arbitrary width. Our findings quantitatively describe the anharmonic Bloch dynamics observed in recent experiments and are applicable in the full crossover regime between Bloch breathing of single site 
excitations and Bloch oscillations of wide wave packets. \\
A promising framework for the experimental realization of cold atom helix lattices is provided by nanofiber-based optical trapping techniques as recently demonstrated \cite{Vetsch2010,Vetsch2012}.
In these experiments, cold atoms are trapped in the two-color evanescent light field above the surface of a tapered optical nanofiber. 
This setup is highly versatile, allowing for combinations of running or standing light waves of different frequencies and polarizations,
and also for local variations of the fiber radius. It was shown that using circularly polarized light results in a continuous double helix shaped potential of tunable parameters \cite{Reitz2012},
while parallel linear polarization gives rise to a sequence of linearly arranged potential wells of equidistant spacing \cite{Vetsch2010}.
These two limiting case can be continuously transformed into each other using elliptically polarized light of varying ellipticity.
In the intermediate regime, this will induce the desired lattice-type modulation of the double helix curve.
In its basic implementation, this scheme will produce two potential wells per winding for each of the helix strands, 
but the number can be increased by quickly switching the orientation of the polarization axes, resulting in a time-averaged potential for the atoms.
\\
In the present work, we have restricted to the single particle dynamics of the helix lattice model. Combining the long range hoppings with a local on-site interaction term will open a variety of directions for future studies. 
Within the bosonic mean field framework, one is led to a discrete nonlinear Schrödinger equation with long range hoppings, reducing to Eq.~(\ref{eq:DS}) in the non-interacting limit. Previous studies have shown that the presence of beyond nearest neighbor hoppings in such a model may drastically alter the properties of its localized excitations, see e.g. \cite{Gaididei1997,Efremidis2002,Szameit2009}, but these considerations have been restricted to second neighbor hopping only or to hoppings decreasing monotonically with the index difference. 
Going beyond the mean field approximation, it has been argued that long range hoppings in a one-dimensional model can effectively mimic higher dimensionalities, see \cite{Tezuka2014} for a recent example.
A helix lattice setup may thus serve to experimentally address the crossover between a purely one-dimensional system (by suppressing inter-winding tunneling) and a system with two-dimensional characteristics (by using many sites per winding and sizable intra- and inter-winding tunnelings). On the theoretical side, such a dimensional crossover is subject to active research even for cubic lattices \cite{Schonmeier-Kromer2014}.
Furthermore, the existence of frustration effects for negative hopping parameters has already been mentioned in Sec.~\ref{sect3}, pointing to interesting ground state features in this regime. 
Notably, the well-known mapping from hard-core bosons to non-interacting fermions often employed for the standard Hubbard model breaks down in the presence of long range hoppings, see e.g. \cite{Corboz2010}.
Finally, we mention the additional possibility of introducing long range interactions into the helix lattice framework, for instance of the dipolar or Coulomb type. 
Recently, such systems with the particle dynamics being constrained to low dimensional structures but the interaction exploring the surrounding space have received considerable attention \cite{Lahaye2009,Law2008,Schmelcher2011,Gammelmark2013,Pedersen2014}.

\begin{acknowledgments}
 We thank A. Rauschenbeutel and P. G. Kevrekidis for insightful discussions.
J. S. gratefully acknowledges support from the {\it Studienstiftung des deutschen Volkes}.
\end{acknowledgments}

\appendix

\begin{widetext}

\section{Double helix lattice model and its reduction}	
\label{app:doublehelix}
The recent proposals \cite{Bhattacharya2007,Reitz2012} for the experimental creation of helical waveguides for ultracold atoms result in double-helix structures, rather than in a single strand helix as discussed here.
In this appendix, we introduce a double helix lattice system and show that its single particle dynamics can be reduced to that of the single strand model.\\
Adopting notation from the main text, we label the two strands of the double helix by (I) and (II) and take them to be parametrized by $\vec{r}^\text{ (I)} (\varphi) = \left(R \cos \varphi,R \sin \varphi, b \varphi \right)$ 
and $\vec{r}^\text{ (II)} (\varphi) = \left(-R \cos \varphi,-R \sin \varphi, b \varphi \right)$, respectively.
We assume a symmetric geometry in which the potential wells forming the lattice sites are spaced equidistantly (in terms of arc-length) on the two strands, with the same site-to-site distance for both strands. 
Furthermore, we take the geometry to be chosen such that lattice sites on the different strands are located at identical $z$-coordinates.
Summarizing, we assume site positions
\begin{equation*}
 \vec r^\text{ (I)}_l = \left(R \cos(l \varphi_0), R \sin( l \varphi_0), b \varphi_0 l\right), \quad  \vec r^\text{ (II)}_l = \left(-R \cos(l \varphi_0), -R \sin( l \varphi_0), b \varphi_0 l\right)
\end{equation*}
for some fixed $\varphi_0$. This makes sure that the Euclidean distances between a site and its $n$-th neighbor on both the same strand and the opposite strand depend on the index difference only.
With Hubbard parameters given by $t_\alpha$ ($\alpha \geq 0$, $t_0 = 0$) for hopping to the $\alpha$-th neighbor in index on the same strand and $t'_\alpha$ ($\alpha \geq 0$) for hopping to the $\alpha$-th neighbor in index on the opposite strand, considerations analogous to those in the main text give rise to two coupled discrete Schrödinger equations for the single particle dynamics under a constant force along the $z$-axis
\begin{eqnarray*}
  \ii \partial_\tau \Psi^\text{(I)}_l &=& - \sum_{\alpha}\left[ t_\alpha \left( \Psi^\text{(I)}_{l+\alpha} + \Psi^\text{(I)}_{l-\alpha} \right) + t'_\alpha \left( \Psi^\text{(II)}_{l+\alpha} + \Psi^\text{(II)}_{l-\alpha} \right) \right] + F l \Psi^\text{(I)}_l , \\
  \ii \partial_\tau \Psi^\text{(II)}_l &=& - \sum_{\alpha}\left[ t_\alpha \left( \Psi^\text{(II)}_{l+\alpha} + \Psi^\text{(II)}_{l-\alpha} \right) + t'_\alpha \left( \Psi^\text{(I)}_{l+\alpha} + \Psi^\text{(I)}_{l-\alpha} \right) \right] + F l \Psi^\text{(II)}_l .
\end{eqnarray*}
This system can be readily decoupled, cf. \cite{Dauxois1991}, via the transformation $\Psi_l^{(\pm)}:= \left( \Psi^\text{(I)}_l \pm \Psi^\text{(II)}_l \right)/ \sqrt{2}$, resulting in
\begin{eqnarray*}
  \ii \partial_\tau \Psi^{(+)}_l &=& - \sum_{\alpha=0}^\infty \left( t_\alpha + t'_\alpha \right) \left( \Psi^{(+)}_{l+\alpha} + \Psi^{(+)}_{l-\alpha} \right) + F l \Psi^{(+)}_l , \\
  \ii \partial_\tau \Psi^{(-)}_l &=& - \sum_{\alpha=0}^\infty \left( t_\alpha - t'_\alpha \right) \left( \Psi^{(-)}_{l+\alpha} + \Psi^{(-)}_{l-\alpha} \right) + F l \Psi^{(-)}_l ,
\end{eqnarray*}
which is a set of two independent copies of the single strand lattice model studied in this work. Thus, all our results can immediately be carried over to the non-interacting double strand lattice. In particular, if the initial wave function is such that opposing sites on the two strands are equally populated and in phase, $\Psi_l^{(-)}$ will vanish for all times, and $\Psi_l^{(+)}=\sqrt{2}\Psi^\text{(I)}_l=\sqrt{2}\Psi^\text{(II)}_l$ will immediately be governed by a discrete Schrödinger equation as Eq.~(\ref{eq:DS}). 

\section{General expressions for the first two moments of the position operator}
\label{app:prop}
In this appendix, we extend the results of Sec.~\ref{sect5} by considering an arbitrary initial wave function $\Psi_l(0)$. We derive closed expressions for the
time dependence of the first two moments of the position operator in terms of this initial condition. 
The general result is then specialized to the one for the Gaussian wave packet as provided in Sec.~\ref{sect5}.\\
Using the explicit form of the propagator given in Eq.~(\ref{eq:prop}), we find for any power $p$ of the position operator
\begin{eqnarray*}
 \langle X^p \rangle_\tau &=& \sum_l l^p \Psi_l^* (\tau) \Psi_l (\tau) = \sum_{l,l',l''} l^p U_{l,l'}(\tau) U^*_{l,l''}(\tau) \Psi_{l'}(0) \Psi^*_{l''}(0) \\
&=& \sum_{l',l''} \Psi_{l'}(0) \Psi^*_{l''}(0) \ii^{l''-l'} e^{\ii \frac{F\tau}{2}(l''-l')} \sum_{ \{ n_\alpha, n'_\alpha \}_{\alpha=2}^A } \ii^{\sum_{\alpha=2}^A (\alpha-1)(n'_\alpha-n_\alpha)} \prod_{\alpha=2}^A J_{n_\alpha}(\xi_\alpha) J_{n'_\alpha}(\xi_\alpha) \\
&&\quad \times \sum_l l^p J_{l-l'-\sum_{\alpha=2}^A \alpha n_\alpha}(\xi_1)J_{l-l''-\sum_{\alpha=2}^A \alpha n'_\alpha}(\xi_1)
\end{eqnarray*}
with the shorthand $\xi_\alpha=\frac{4 t_\alpha}{\alpha F} \sin \frac{\alpha F \tau }{2}$. The strategy is now to employ the recursion and orthogonality relations of the Bessel functions provided in Eq.~(\ref{eq:bessel2}) to first evaluate the sum over $l$.
We show this explicitly for $p=1$, where Eq.~(\ref{eq:bessel2}) yields
\begin{eqnarray*}
&& \sum_l l J_{l-l'-\sum_{\alpha=2}^A \alpha n_\alpha}(\xi_1)J_{l-l''-\sum_{\alpha=2}^A \alpha n'_\alpha}(\xi_1)\\
 &=& \left(l' + \sum_{\beta=2}^A \beta n_\beta \right) \delta_{l'' -l', \sum_{\alpha=2}^A \alpha (n_\alpha- n'_\alpha)}
+ \frac{\xi_1}{2} \left( \delta_{l'' -l', \sum_{\alpha=2}^A \alpha (n_\alpha- n'_\alpha)+1} + \delta_{l'' -l', \sum_{\alpha=2}^A \alpha (n_\alpha- n'_\alpha)-1} \right).
\end{eqnarray*}
Inserting this, evaluating the sum over $l''$ using the Kronecker symbols and introducing the summation indices $m_\alpha= n_\alpha-n'_\alpha$, this results in
\begin{eqnarray*}
 \langle X \rangle_\tau &=& \sum_{l'} \sum_{ \{ n_\alpha, m_\alpha \} } \Psi_{l'}(0) \Psi^*_{l'+\sum_\alpha \alpha m_\alpha}(0) \ii^{\sum_\alpha m_\alpha} e^{\ii \frac{F \tau}{2} \sum_\alpha \alpha m_\alpha } \left(l'+ \sum_\beta \beta n_\beta \right) \prod_\alpha J_{n_\alpha}(\xi_\alpha) J_{n_\alpha - m_\alpha}(\xi_\alpha) \\
&& + \frac{\xi_1}{2} \ii e^{\ii \frac{F \tau}{2}} \sum_{l'} \sum_{ \{ n_\alpha, m_\alpha \} } \Psi_{l'}(0) \Psi^*_{l'+\sum_\alpha \alpha m_\alpha+1}(0) \ii^{\sum_\alpha m_\alpha} e^{\ii \frac{F \tau}{2} \sum_\alpha \alpha m_\alpha }  \prod_\alpha J_{n_\alpha}(\xi_\alpha) J_{n_\alpha - m_\alpha}(\xi_\alpha) \\
&& - \frac{\xi_1}{2} \ii e^{-\ii \frac{F \tau}{2}} \sum_{l'} \sum_{ \{ n_\alpha, m_\alpha \} } \Psi_{l'}(0) \Psi^*_{l'+\sum_\alpha \alpha m_\alpha-1}(0) \ii^{\sum_\alpha m_\alpha} e^{\ii \frac{F \tau}{2} \sum_\alpha \alpha m_\alpha }  \prod_\alpha J_{n_\alpha}(\xi_\alpha) J_{n_\alpha - m_\alpha}(\xi_\alpha),
\end{eqnarray*}
where all indices $\alpha, \beta$ range from $2$ to $A$. In the next step, the recursion and orthogonality relations of Eq.~(\ref{eq:bessel2}) can be employed again to evaluate the sums over $n_\alpha$ and $m_\alpha$ for each $\alpha$, which finally gives
\begin{equation}
 \langle X \rangle_\tau = \sum_l l |\Psi_l(0)|^2 + \sum_{\alpha=1}^A \alpha \xi_\alpha \text{Re} \left[ \ii e^{\ii \frac{\alpha F \tau }{2}} \sum_l \Psi_l(0) \Psi_{l+\alpha}^*(0) \right]
= \langle X \rangle_0 + \sum_{\alpha=1}^A \frac{2 t_\alpha}{F} \text{Re} \left[ \left( e^{\ii \alpha F \tau } -1 \right) g(\alpha) \right],
\label{eq:xgen}
\end{equation}
where we have introduced $g(\alpha):=\sum_l \Psi_l(0) \Psi_{l+\alpha}^*(0)$, see also the recent finding within the nearest neighbor model \cite{Dominguez-Adame2010}.
The calculation of $\langle X^2 \rangle_\tau$ is more involved but proceeds along the same lines, so we only state the final result:
\begin{eqnarray}
\langle X^2 \rangle_\tau &=& \langle X^2 \rangle_0 + \sum_{\alpha=1}^A 8 \frac{t^2_\alpha}{F^2} \sin^2 \frac{\alpha F \tau }{2} \left( 1 - \text{Re} \left[ e^{\ii \alpha F \tau } g(2\alpha) \right] \right)
+ \sum_{\alpha=1}^A \frac{2 t_\alpha}{F} \text{Re} \left[ \left(e^{\ii \alpha F \tau } -1 \right) \sum_l (2l+\alpha) \Psi_l(0) \Psi_{l+\alpha}^*(0) \right] \nonumber \\
&& \quad + {\sum_{\alpha, \beta=1, \alpha \neq \beta}^A }8 \frac{t_\alpha t_\beta}{F^2} \sin \frac{\alpha F \tau }{2}\sin \frac{ \beta F \tau}{2} \text{Re} \left[ e^{\ii (\alpha - \beta) \frac{F \tau}{2}} g(\alpha- \beta) - e^{\ii (\alpha + \beta) 
\frac{F \tau}{2}} g(\alpha + \beta) \right].  
\label{eq:x2gen}
\end{eqnarray}
Now if we take the initial wave function to be real and symmetric with respect to a site $l_0$, i.e. $\Psi_{l_0+l}(0) = \Psi_{l_0 - l}(0)$, it can be seen that $\sum_l (2l+\alpha) \Psi_l(0) \Psi_{l+\alpha}(0)= 2 \langle X \rangle_0 g(\alpha)$, yielding the simplified expressions
\begin{eqnarray*}
 \langle X \rangle_\tau &=& \langle X \rangle_0 - \sum_{\alpha=1}^A \frac{4 t_\alpha}{F}  g(\alpha) \sin^2 \frac{  \alpha F \tau}{2} , \\
 \langle X^2 \rangle_\tau &=& \langle X^2 \rangle_0 + \sum_{\alpha=1}^A 8 \frac{t^2_\alpha}{F^2} \sin^2 \frac{\alpha F \tau }{2} \left[ 1 - g(2\alpha) \cos (\alpha F \tau)  \right]
- \langle X \rangle_0 \sum_{\alpha=1}^A \frac{8 t_\alpha}{F} g(\alpha) \sin^2 \frac{\alpha F \tau }{2} \nonumber \\
&& \quad + {\sum_{\alpha, \beta=1, \alpha \neq \beta}^A }8 \frac{t_\alpha t_\beta}{F^2} \sin \frac{\alpha F \tau }{2}\sin \frac{\beta F \tau }{2} \left[ g(\alpha- \beta) \cos \frac{(\alpha - \beta)F \tau}{2} 
- g(\alpha + \beta) \cos \frac{(\alpha + \beta)F \tau}{2}  \right]. 
\end{eqnarray*}
Specializing to the Gaussian wave packet centered at site $l_0 =0$, we furthermore have $\langle X \rangle_0 = 0$ and the functions $g(\alpha)$ can be evaluated in terms of Theta functions using the sum \cite{Bellmann1961}
\begin{equation*}
 \sum_l e^{-\left(l^2 + (l+\alpha)^2 \right)/(2 \sigma)} = e^{-\alpha^2/(4\sigma)} \sqrt{\sigma \pi} \,\, \vartheta_3 \left( \alpha \frac{\pi}{2}, e^{-\sigma \pi^2} \right).
\end{equation*}
This results in Eqs.~(\ref{eq:x},\ref{eq:x2}) of the main text.

\end{widetext}

\bibliography{bibliography}

\end{document}